%% file: main.tex
\documentclass{article}
\usepackage{times}
\usepackage{eepic,epic}
\usepackage{latexsym}
\usepackage[latin1]{inputenc}    

\makeatletter%
\def\nottoobig#1{{\hbox{$\left#1\vcenter to1.111\ht\strutbox{}\right.\n@space$}}}
\makeatother%

\makeatother%

\makeatletter \@beginparpenalty=10000 \makeatother

\def\underl#1 {\leavevmode\let\first=\relax\underli #1 }
\def\underli#1 {\ifx&#1\let\next=\relax\unskip
                \else\let\next=\underli\first\ulinebox{#1}\fi\let\first=\undersp\next}
\def\undersp{\penalty50\ulinebox{\space}\penalty50}
\def\ulinebox#1{\vtop{\hbox{\strut#1}\hrule}}%
\def\unice#1 {\underl #1 & }
\def\desclabel#1{\bf #1\hfil}
\def\desc{\list{}{%
\labelwidth=\leftmargin
\advance \labelwidth by -\labelsep
\let \makelabel=\desclabel}}

\makeatletter %

\newlength{\filength}
\newsavebox{\gcbox}
\sbox{\gcbox}{\framebox[\filength]{\rule{0ex}{2ex}}}

\newlength{\leftjustindent}
\newlength{\@leftjustindent}
\def\leftjust{\let\\\@leftjustcr\let\end\@endleftjust
  \addtolength{\@leftjustindent}{\leftjustindent}
  \vcenter\bgroup
  \halign\bgroup
    \hbox to\displaywidth{
      \rule{\@leftjustindent}{0ex}$\displaystyle##$\hfill
      }\crcr
}
\def\endleftjust{\crcr\egroup\egroup\endgroup}
\def\@endleftjust#1{\crcr\egroup\egroup\@checkend{#1}\endgroup}
\def\@leftjustcr{\crcr}

\newtheorem{theorem}{Theorem}%

\newenvironment{proofs}{\noindent{\bf Proof.}\hspace*{1em}}{\qed\bigskip}

\newcommand{\sproofof}[1]{\noindent{\bf Proof of {#1}.}\hspace*{1em}}
\newcommand{\eproofof}[1]{\hfill \mbox{\qed$_{\mbox{\small {#1}}}$}\quad\medskip}

\newcommand{\qedblob}{\mbox{\rule[-1.5pt]{5pt}{10.5pt}}}
\def\literalqed{{\ \nolinebreak\hfill\mbox{\qedblob\quad}}}

\def\qed{\literalqed}

\newtheorem{lemma}[theorem]{Lemma}

\newtheorem{proposition}[theorem]{Proposition}
\newcommand{\singlespacing}{\let\CS=
\@currsize\renewcommand{\baselinestretch}{1}\tiny\CS}
\newcommand{\singlespacingplus}{\let\CS=
\@currsize\renewcommand{\baselinestretch}{1.25}\tiny\CS}
\newcommand{\doublespacing}{\let\CS=
\@currsize\renewcommand{\baselinestretch}{1.75}\tiny\CS}
\newcommand{\draftspacing}{\let\CS=
\@currsize\renewcommand{\baselinestretch}{2.0}\tiny\CS}
\newcommand{\foospacing}{\let\CS=
\@currsize\renewcommand{\baselinestretch}{1.05}\tiny\CS}

\makeatother%

\mathcode`\0="0030      %
\mathcode`\1="0031
\mathcode`\2="0032
\mathcode`\3="0033
\mathcode`\4="0034
\mathcode`\5="0035
\mathcode`\6="0036
\mathcode`\7="0037
\mathcode`\8="0038
\mathcode`\9="0039

\newtheorem{definition}[theorem]{Definition}

\makeatletter
\clubpenalty=\@highpenalty
\widowpenalty=\@highpenalty
\makeatother

\makeatletter
\newcommand{\niceonespacing}{\let\CS=\@currsize\renewcommand{\baselinestretch}{1.1}\tiny\CS}\newcommand{\nicetwospacing}{\let\CS=\@currsize\renewcommand{\baselinestretch}{1.2}\tiny\CS}
\newcommand{\nicethreespacing}{\let\CS=\@currsize\renewcommand{\baselinestretch}{1.3}\tiny\CS}
\newcommand{\singlespacingplusplus}{\let\CS=\@currsize\renewcommand{\baselinestretch}{1.35}\tiny\CS}
\newcommand{\nicefourspacing}{\let\CS=\@currsize\renewcommand{\baselinestretch}{1.4}\tiny\CS}
\newcommand{\nicefivespacing}{\let\CS=\@currsize\renewcommand{\baselinestretch}{1.5}\tiny\CS}
\newcommand{\nicesixpacing}{\let\CS=\@currsize\renewcommand{\baselinestretch}{1.6}\tiny\CS}
\makeatother

\makeatletter
\def\@cite#1#2{[#1\if@tempswa , #2\fi]}
\makeatother

\makeatletter
\def\@citex[#1]#2{\if@filesw\immediate\write\@auxout{\string\citation{#2}}\fi
  \def\@citea{}\@cite{\@for\@citeb:=#2\do
    {\@citea\def\@citea{,\linebreak[0]}\@ifundefined
       {b@\@citeb}{{\bf ?}\@warning
       {Citation `\@citeb' on page \thepage \space undefined}}%
\hbox{\csname b@\@citeb\endcsname}}}{#1}}
\makeatother

\makeatletter
\def\ps@thesis{\def\@oddhead{\hfil\rm\thepage\hfil}\def\@oddfoot{}\def\@evenhead{\hfil\rm\thepage\hfil}\def\@evenfoot{}\def\chaptermark##1{}\def\sectionmark##1{}}
\makeatother

\makeatletter
\def\foobarpt{\textfont\z@\tenrm 
  \scriptfont\z@\ninrm \scriptscriptfont\z@\sevrm
\textfont\@ne\tenmi \scriptfont\@ne\ninmi \scriptscriptfont\@ne\sevmi
\textfont\tw@\tensy \scriptfont\tw@\ninsy \scriptscriptfont\tw@\sevsy
\textfont\thr@@\tenex \scriptfont\thr@@\tenex \scriptscriptfont\thr@@\tenex
\def\unboldmath{\everymath{}\everydisplay{}\@nomath\unboldmath
          \textfont\@ne\tenmi 
          \textfont\tw@\tensy \textfont\lyfam\tenly
          \@boldfalse}\@boldfalse
\def\boldmath{\@ifundefined{tenmib}{\global\font\tenmib\@mbi\@magscale1\global
        \font\tensyb\@mbsy \@magscale1\global\font
         \tenlyb\@lasyb\@magscale1\relax\@addfontinfo\@xiipt
              {\def\boldmath{\everymath
                {\mit}\everydisplay{\mit}\@prtct\@nomathbold
                \textfont\@ne\tenmib \textfont\tw@\tensyb 
                \textfont\lyfam\tenlyb\@prtct\@boldtrue}}}{}\@xiipt\boldmath}%
\def\prm{\fam\z@\tenrm}%
\def\pit{\fam\itfam\tenit}\textfont\itfam\tenit \scriptfont\itfam\ninit
   \scriptscriptfont\itfam\sevit
\def\psl{\fam\slfam\tensl}\textfont\slfam\tensl 
     \scriptfont\slfam\tensl \scriptscriptfont\slfam\tensl
\def\pbf{\fam\bffam\tenbf}\textfont\bffam\tenbf 
   \scriptfont\bffam\ninbf \scriptscriptfont\bffam\ninbf 
\def\ptt{\fam\ttfam\tentt}\textfont\ttfam\tentt
   \scriptfont\ttfam\nintt \scriptscriptfont\ttfam\nintt 
\def\psf{\fam\sffam\tensf}\textfont\sffam\tensf
    \scriptfont\sffam\tensf \scriptscriptfont\sffam\tensf
\def\psc{\@getfont\psc\scfam\@xiipt{\@mcsc\@magscale1}}%
\def\ly{\fam\lyfam\tenly}\textfont\lyfam\tenly 
   \scriptfont\lyfam\ninly \scriptscriptfont\lyfam\sevly
 \@setstrut \rm}

\makeatother

\newcommand{\p}{\mbox{\rm P}}

\newcommand{\np}{\mbox{\rm NP}}

\newcommand{\condition}{\,\nottoobig{|}\:}

\newcommand{\openNeighbors}{\mbox{$\mbox{\rm{}openNeighbors}_{\mathcal{P}}$}}
\newcommand{\openSets}{\mbox{$\mbox{\rm{}openSets}_{\mathcal{P}}$}}
\newcommand{\balance}{\mbox{$\mbox{\rm{}balance}_{\mathcal{P}}$}}
\newcommand{\surplus}{\mbox{$\mbox{\rm{}surplus}_{\mathcal{P}}$}}
\newcommand{\degree}{\mbox{\it deg}}
\newcommand{\area}{\mbox{$\mbox{\rm{}area}_{\mathcal{P}}$}}
\newcommand{\gap}{\mbox{$\mbox{\rm{}gap}_{\mathcal{P},\mathcal{A}}$}}
\newcommand{\sumgap}{\mbox{$\mbox{\rm{}sumgap}_{\mathcal{P},\mathcal{A}}$}}
\newcommand{\maxgap}{\mbox{$\mbox{\rm{}maxgap}_{\mathcal{P},\mathcal{A}}$}}
\newcommand{\mingap}{\mbox{$\mbox{\rm{}mingap}_{\mathcal{P},\mathcal{A}}$}}
\newcommand{\maxdegree}{\mbox{\it max-deg}}
\newcommand{\mindegree}{\mbox{\it min-deg}}

\newcommand\seq{\subseteq}

\newcommand{\bigoh}{\mathcal{O}}

\newenvironment{construction}{\bigbreak\begin{block}}{\end{block}
    \bigbreak}

\newenvironment{block}{\begin{list}{\hbox{}}{\leftmargin 1em
    \itemindent -1em \topsep 0pt \itemsep 0pt \partopsep 0pt}}{\end{list}}

\newcommand{\dominate}{{\mbox{\sc Dominate}}}
\newcommand{\assign}{{\mbox{\sc Assign}}}
\newcommand{\recalculateGaps}{{\mbox{\sc Recalculate-Gaps}}}

\newcommand{\handleCriticalVertex}{{\mbox{\sc Handle-Critical-Vertex}}}

\newcounter{alg}

\title{An Exact $2.9416^n$ Algorithm for the Three Domatic Number
  Problem\thanks{Work supported in part by the DFG under Grant RO~1202/9-1.}}

\author{
Tobias Riege\thanks{Email: ${\tt riege@cs.uni\mbox{-}duesseldorf.de}$.}
\quad and \quad
J\"{o}rg Rothe\thanks{Email: ${\tt rothe@cs.uni\mbox{-}duesseldorf.de}$.} \\
Institut f\"ur Informatik \\
Heinrich-Heine-Universit\"at D\"usseldorf \\
40225 D\"usseldorf, Germany
}

\date{June 24, 2005}

\makeatletter
\def\@listI{\leftmargin\leftmargini \parsep 4.5pt plus 1pt minus 1pt\topsep
6pt plus 2pt minus 2pt \itemsep  2pt plus 2pt minus 1pt}

\let\@listi\@listI
\@listi
\makeatother

\begin{document}

\typeout{WARNING:  BADNESS used to suppress reporting!  Beware!!}
\hbadness=3000%
\vbadness=10000 %

\setcounter{page}{1}

\sloppy

\setcounter{footnote}{0}

\maketitle

\begin{abstract}
\noindent
The three domatic number problem asks whether a given undirected graph
can be partitioned into at least three dominating sets, i.e., sets
whose closed neighborhood equals the vertex set of the graph.  Since
this problem is NP-complete, no polynomial-time algorithm is known for
it.  The naive deterministic algorithm for this problem runs in
time~$3^n$, up to polynomial factors.  In this paper, we design an
exact deterministic algorithm for this problem running in
time~$2.9416^n$.  Thus, our algorithm can handle problem instances of
larger size than the naive algorithm in the same amount of time.  We
also present another deterministic and a randomized algorithm for this
problem that both have an even better performance for graphs with
small maximum degree.

\vspace*{.5cm}
\noindent
\begin{tabular}{ll}
{\bf Key words:}
 & Exact algorithms, domatic number problem 
\end{tabular}
\end{abstract}

\setcounter{page}{1}
\pagestyle{plain}
\sloppy

\setcounter{page}{1}

\section{Introduction}

In this paper, we design a deterministic algorithm for the three
domatic number problem, which is one of the standard $\np$-complete
problems, see Garey and Johnson~\cite{gar-joh:b:int}.  This problem
asks, given an undirected graph~$G$, whether or not the vertex set of
$G$ can be partitioned into three dominating sets.  A dominating set
is a subset of the vertex set that ``dominates'' the graph in that its
closed neighborhood covers the entire graph.  Motivated by the tasks
of distributing resources in a computer network and of locating
facilities in a communication network, this problem and the related
problem of finding a minimum dominating set in a given graph have been
thoroughly studied, see, e.g.,
\cite{coc-hed:j:towards-theory-of-domination,far:j:domatic-number-strongly-chordal-graphs,bon:j:domatic-number-circular-arc-graphs,kap-sha:j:domatic-number,heg-tel:j:generalized-dominating-sets,fei-hal-kor:c:approximating-domatic-number,rie-rot:j:exact-dnp}.

The exact (i.e., deterministic) algorithm designed in this paper runs
in exponential time.  However, its running time is better than that of
the naive exact algorithm for this problem.  That is, we improve the
trivial $\tilde{\mathcal{O}}(3^n)$ time bound to a time bound
of~$\tilde{\mathcal{O}}(2.9416^n)$, where the $\tilde{\mathcal{O}}$
notation neglects polynomial factors as is common for exponential-time
algorithms.  The point of such an improvement is that a
$\tilde{\mathcal{O}}(c^n)$ algorithm, where $c < 3$ is a constant, can
deal with larger instances than the trivial $\tilde{\mathcal{O}}(3^n)$
algorithm in the same amount of time before the exponential growth
rate eventually hits and the running time becomes infeasible.  For
example, if $c = \sqrt{3} \approx 1.732$ then we have
$\tilde{\bigoh}\left(\sqrt{3}^{\, 2n}\right) = \tilde{\bigoh}(3^n)$,
so one can deal with inputs twice as large as before.  Doubling the
size of inputs that can be handled by some algorithm can make quite a
difference in practice.

Exact exponential-time algorithms with improved running times have
been designed for various other important $\np$-complete problems.
For example, Dantsin et al.~\cite{dan-etal:j:det-alg-for-ksat} pushed
the trivial $\tilde{\bigoh}(2^n)$ bound for the three satisfiability
problem down to~$\tilde{\bigoh}(1.481^n)$, which was further improved
to $\tilde{\bigoh}(1.473^n)$ by Brueggemann and
Kern~\cite{bru-ker:t:local-search-threesat}.
Schöning~\cite{sch:j:constraint-satisfaction}, Hofmeister et
al.~\cite{hof-sch-sch-wat:c:probabilistic-threesat-alg-further-improved}
and Paturi et al.~\cite{pat-pud-sak-zan:c:exptime-algorithm-for-ksat}
proposed even better randomized algorithms for the satisfiability
problem.  Combining their ideas, the currently best randomized
algorithm for this problem is due to Iwama and
Tamaki~\cite{iwa-tak:t:improved-upper-bound-threesat}, who achieve a
time bound of~$\tilde{\bigoh}(1.324^n)$.

The currently best exact time bound of $\tilde{\bigoh}(1.211^n)$ for
the independent set problem is due to
Robson~\cite{rob:j:exact-alg-for-independent-set}.
Eppstein~\cite{epp:c:algs-for-three-sat-threecolor,epp:c:exact-alg-for-graph-coloring}
achieved a $\tilde{\bigoh}(2.415^n)$ time bound for graph coloring and
a $\tilde{\bigoh}(1.3289^n)$ for the special case of graph three
colorability.  Fomin, Kratsch, and
Woeginger~\cite{fom-kra-woe:c:exact-algorithm-for-dominating-set}
improved the trivial $\tilde{\bigoh}(2^n)$ bound for the dominating
set problem to~$\tilde{\bigoh}(1.93782^n)$.  Comprehensive surveys on
this subject have been written by
Woeginger~\cite{woe:b:exact-algorithms-for-np-hard-problems} and
Schöning~\cite{sch:c:algorithmics-in-exponential-time}.

In designing domatic number algorithms, it might be tempting to exploit known
results (such as Eppstein's $\tilde{\bigoh}(1.3289^n)$ bound) for the
graph three colorability problem, which resembles the three domatic
number problem in that both are partitioning problems.  However, as
Cockayne and Hedetniemi~\cite{coc-hed:j:towards-theory-of-domination}
point out, the theory of domination is dual to the theory of
coloring in the following sense.  Coloring is based on the hereditary
property of independence.  A graph property
is {\em hereditary\/} if whenever some set of vertices has the
property then so does every subset of it.  In contrast, domination is
an {\em expanding\/} property in that every superset of a dominating
set also is a dominating set of the graph.  Further, graph colorability is a
minimum problem, whereas the domatic 
number problem is a maximum problem.  Independence (and thus
colorability) can be seen as a {\em local\/} property, since it
suffices to check the immediate neighborhood of a set of vertices to
determine whether or not it is independent.  In contrast, dominance is
a {\em global\/} property, since in order to check it one has to
consider the relation between the given set of vertices and the entire
graph.  In this sense, determining the domatic number of a graph
intuitively appears to be harder than computing its chromatic number,
notwithstanding that both problems are $\np$-complete.  More to the
point, the algorithms developed for graph coloring seem to be of no
help in designing algorithms for dominating set or domatic number
problems.

After introducing some definitions and notation in
Section~\ref{sec:prelims}, we describe and analyze our algorithm in
Section~\ref{sec:algo}; the actual pseudo-code is shifted to the
appendix.  In Section~\ref{sec:bounded-max-degree}, we give another
deterministic and a randomized algorithm, which have an even better
running time for graphs with small maximum degree. Finally, we
summarize and discuss our results in Section~\ref{sec:conclusion}.

\section{Preliminaries and Simple Observations}
\label{sec:prelims}

We start by introducing some graph-theoretical notation.  We only
consider simple, undirected graphs without loops in this paper.  Let
$G = (V,E)$ be a graph.  Unless stated otherwise, $n$ denotes the
number of vertices in~$G$. The {\em neighborhood of a vertex~$v$ in
$V$\/} is defined by $N(v) = \{ u \in V \condition \{u,v\} \in E \}$,
and the {\em closed neighborhood of~$v$\/} is defined by $N[v] = N(v)
\cup \{ v \}$.  For any subset $S \seq V$ of the vertices of~$G$,
define $N[S] = \bigcup_{v \in S} N[v]$ and $N(S) = N[S] - S$.  The
{\em degree of a vertex $v$ in $G$\/} is the number of vertices
adjacent to~$v$, i.e., $\degree_G(v) = ||N(v)||$.  If the graph $G$ is
clear from the context, we omit the subscript~$G$.  Define the {\em
minimum degree in $G$\/} by $\mindegree(G) = \min_{v \in V}
\degree(v)$, and the {\em maximum degree in $G$\/} by $\maxdegree(G) =
\max_{v \in V} \degree(v)$.  
A path $P_k = u_1 u_2 \cdots u_k$ of length 
$k$ is a sequence of $k$ vertices, where each vertex is adjacent to
its successor, i.e., $\{u_i, u_{i+1} \} \in E$ for $1 \leq i \leq k-1$. 
If, in addition, $\{ u_k, u_1 \} \in E$, then path $P_k$ is said
to be a {\em cycle}, and we write $C_k$ instead of $P_k$.

\begin{definition}  
Let $G = (V,E)$ be a graph.  A subset $D \seq V$ is a {\em dominating
set of~$G$\/} if and only if $N[D] = V$, i.e., if and only if every
vertex in $G$ either belongs to~$D$ or has some neighbor in~$D$.  The
{\em domination number of~$G$}, denoted~$\gamma(G)$, is the minimum
size of a dominating set of~$G$. The {\em domatic number of~$G$},
denoted~$\delta(G)$, is the maximum number of disjoint dominating sets
of~$G$, i.e., $\delta(G)$ is the maximum $k$ such that $V = V_1 \cup
V_2 \cup \ldots \cup V_k$, where $V_i \cap V_j = \emptyset$ for $1
\leq i < j \leq k$, and each $V_i$ is a dominating set of~$G$.
The {\em dominating set problem\/} asks, given a graph $G$ and a
positive integer~$k$, whether or not $\gamma(G) \leq k$.
The {\em domatic number problem\/} asks, given a graph $G$ and a
positive integer~$k$, whether or not $\delta(G) \geq k$.
\end{definition}

For fixed $k \geq 3$, both the dominating set problem and the domatic
number problem are known to be $\np$-complete, see Garey and
Johnson~\cite{gar-joh:b:int}.  Thus, they are not solvable in
deterministic polynomial time unless $\p = \np$, and all we can hope
for is to design an exponential-time algorithm having a better running
time than the trivial exponential time bound.
For exponential-time algorithms, it is common to drop polynomial factors,
as indicated by the $\tilde{\bigoh}$ notation: For functions $f$
and~$g$, we write $f \in \tilde{\bigoh}(g)$ if and only if $f \in
\bigoh(p \cdot g)$ for some polynomial~$p$.
The naive deterministic
algorithm for the dominating set problem runs in time
$\tilde{\bigoh}(2^n)$.  Fomin, Kratsch, and
Woeginger~\cite{fom-kra-woe:c:exact-algorithm-for-dominating-set}
improved this trivial upper bound to~$\tilde{\bigoh}(1.93782^n)$.  For
various restricted graph classes, they achieve even better bounds.

The naive deterministic algorithm for the domatic number problem works
as follows: Given a graph~$G$ and an integer~$k$, it sequentially
checks every potential solution (i.e., every possible partition of the
vertex set of $G$ into $k$ sets $D_1, D_2, \ldots, D_k$), and accepts
if and only if a correct solution is found (i.e., if and only if each
$D_i$ is a dominating set).  How many potential solutions are there?
The number of ways of partitioning a set with $n$ elements into $k$
nonempty, disjoint subsets
can be calculated by the Stirling number of the second kind:
$S_2(n,k) = \frac{1}{k!} \sum_{i=0}^{k-1} (-1)^i {k \choose i} (k-i)^n$,
which yields a running time of~$\tilde{\bigoh}(k^n)$.
A better result can be achieved via the dynamic programming across the subsets
technique, which was introduced by Lawler~\cite{law:j:complexity-chromatic} to
compute the chromatic number of a graph by exploiting the fact that every
minimum chromatic partition contains at least one maximum independent set.  By
suitably modifying this technique, one can compute the domatic number of a
graph in time~$\tilde{\bigoh}(3^n)$. This is done by generating all
dominating sets of the graph with increasing cardinality, which takes time
\[
\sum_{k = 0}^n {n \choose k} 2^k = (1 + 2)^n = 3^n.
\]
The difference to Lawler's algorithm lies in the fact that all
dominating sets need to be checked, whereas only maximum independent
sets are relevant to compute the chromatic number.

\begin{proposition}
\label{prop:lawler-domatic}
Let $G = (V,E)$ be a graph. Then, the domatic number $\delta(G)$ can
be computed in time~$\tilde{\bigoh}(3^n)$.
\end{proposition}

One tempting way of designing an improved algorithm for the domatic number
problem might be to exploit the result for the dominating set problem
mentioned above.  However, we observe that no such useful connection between
the two problems exists in general.
The first part of Proposition~\ref{prop:mds-dnp} shows that an
arbitrary given minimum dominating set is not necessarily part of a
partition into a maximum number of dominating sets.  The second part
of Proposition~\ref{prop:mds-dnp} shows that, given an arbitrary
partition into a maximum number of dominating sets, it is not
necessarily the case that one set of the partition indeed is a minimum
dominating set.  Thus, for solving the domatic number problem, one
cannot use in any obvious way the exact $\tilde{\bigoh}(1.93782^n)$
algorithm for the dominating set problem by Fomin et
al.~\cite{fom-kra-woe:c:exact-algorithm-for-dominating-set}.
Proposition~\ref{prop:mds-dnp} is stated for graphs with domatic number~$3$;
it can easily be generalized to graphs with domatic number $k \geq 3$.
The proof of Proposition~\ref{prop:mds-dnp} can be found in the appendix.

\begin{proposition}
\label{prop:mds-dnp}
\begin{enumerate}
\item There exists some graph $G$ with $\delta(G) = 3$ such that some minimum
  dominating set $D$ of $G$ is not part of any partition into three dominating
  sets of~$G$.
  
\item There exists some graph $H = (V,E)$ with $\delta(H) = 3$ such
  that for each partition $V = D_1 \cup D_2 \cup D_3$ into three
  dominating sets of $H$ and for each~$i$, $||D_i|| > \gamma(H)$.
\end{enumerate}
\end{proposition}

For the three domatic number problem, no algorithm with a running
time better than~$\tilde{\bigoh}(3^n)$ is known.
We improve this trivial upper bound to~$\tilde{\bigoh}(2.9416^n)$.

We now define some technical notions suitable to measure how
``useful'' a vertex is to achieve domination of the graph~$G = (V,E)$.
Intuitively, the vertex degree is a good (local) measure, since the
larger the neighborhood of a vertex is, the more vertices are
potentially dominated by the set to which it belongs.  The technical
notions introduced in Definition~\ref{def:notions-for-algorithm} will
be used later on to describe our algorithm.

\begin{definition}
\label{def:notions-for-algorithm}
  Let $G = (V,E)$ be a graph with $n$ vertices, and let $\mathcal{P} =
  (D_1, D_2, D_3, R)$ be a partition of $V$ into four sets, $D_1$,
  $D_2$, $D_3$, and~$R$.  The subsets $D_i$ of $V$ will eventually
  yield a partition of $V$ into the three dominating sets (if they
  exist) to be constructed, and the subset $R \seq V$ collects the
  remaining vertices not yet assigned at the current point in the
  computation of the algorithm.  Let $r = || R ||$ be the number of
  these remaining vertices, and let $d = n-r$
be the number of vertices already assigned to some set~$D_i$. The {\em
  area of $G$ covered by $\mathcal{P}$} is defined as
  $\area(G) = \sum_{i=1}^3 || N[D_i] ||$.
Note that $\area(G) = 3n$ if and only if $D_1$, $D_2$, and $D_3$ are
dominating sets of~$G$. For a partition~$\mathcal{P}$, we also define
the {\em surplus of graph $G$} as
$\surplus(G) = \area(G) - 3d$.
 
Some of the vertices in $R$ may be assigned to three, not necessarily
disjoint, auxiliary sets~$A_1$, $A_2$, and~$A_3$ arbitrarily.  Let
$\mathcal{A} = (A_1, A_2, A_3)$.  For each vertex $v \in R$ and for
each $i$ with $1 \leq i \leq 3$, define the {\em gap of vertex $v$
with respect to set $D_i$\/} by
\[
\gap(v,i) = \left\{
\begin{array}{ll}
 || N[v] ||
 - || \{ u \in N[v] \condition (\exists w \in N[u]) [ w \in D_i] \} || 
 & \mbox{if $v \notin A_i$} \\
 \bot & \mbox{otherwise,}
\end{array} 
\right.
\]
where $\bot$ is a special symbol that indicates that $\gap(v,i)$ is
undefined for this $v$ and~$i$.  (Our algorithm will make sure
to properly handle the cases of undefined gaps.)

Additionally, given $\mathcal{P}$ and~$\mathcal{A}$, define
for all vertices $v \in R$:
\begin{eqnarray*}
\maxgap(v)
 & = & \max \{ \gap(v,i) \condition 1 \leq i \leq 3 \}, \\
\mingap(v)
 & = & \min \{ \gap(v,i) \condition 1 \leq i \leq 3 \}, \\
\sumgap(v)
 & = & \sum_{i=1}^{3} \gap(v,i).
\end{eqnarray*}

Given $G$, $\mathcal{P}$, and~$\mathcal{A}$, define the {\em maximum
gap of $G$\/} and the {\em minimum gap of $G$\/} by taking the maximum
and minimum gaps over all vertices in~$G$ not yet assigned:
\begin{eqnarray*}
\maxgap(G)
 & = & \max \{ \maxgap(v) \condition v \in R \}, \\
\mingap(G)
 & = & \min \{ \mingap(v) \condition v \in R \}.
\end{eqnarray*}

Let $\mathcal{P}$ be given. A vertex $u \in V$ is called an
{\em open neighbor of $v \in V$\/} if 
$u \in N[v]$ and $u$ has not been assigned
to any set $D_1$, $D_2$, or $D_3$ yet.  A potential dominating set
$D_i$, $1 \leq i \leq 3$, is called an {\em open set of $v \in V$\/}
if its closed neighborhood does not include~$v$, i.e., $v$ is not
dominated by $D_i$. The {\em balance of $v \in V$} is defined as
the difference between the number of open vertices and the number
of open sets.  Formally, define
\begin{eqnarray*}
\openNeighbors(v) & = & \{ u \in N[v] \condition u \in R \}, \\
\openSets(v) & = & \{ i \in \{1,2,3 \} \condition
 v \notin N[D_i] \}, \\
\balance(v) & = & || \openNeighbors(v) || - || \openSets(v) ||.
\end{eqnarray*}
We call a vertex $v \in V$ {\em critical\/} if and only if
$\balance(v) \leq 0$ and $|| \openSets(v) || > 0$.
\end{definition}

The proof of the next proposition is straightforward. Once
$\balance(v) = 0$, no two vertices remaining in $N[v] \cap R$ can
be assigned to the same dominating set $D_i$, $1 \leq i \leq 3$,
since $\balance(v)$ would then be negative.

\begin{proposition}
\label{prop:critical}
Let $\mathcal{P} = (D_1, D_2, D_3, R)$ be given as in
Definition \ref{def:notions-for-algorithm} , and $v \in V$ be
a critical vertex for this partition. The only way to modify
$\mathcal{P}$ so as to contain three dominating sets is to assign
all vertices $u \in N[v] \cap R$ to distinct dominating sets~$D_i$.
\end{proposition}

\section{The Algorithm}
\label{sec:algo}

Our strategy is to recursively assign the vertices $v \in V$ to obtain
a correct potential solution consisting of a partition into three
dominating sets, $D_1$, $D_2$, and~$D_3$.  Once a previous assignment
of $v$ to some set $D_i$ turns out to be wrong, we remember this by
adding this vertex to~$A_i$.  More precisely, the basic idea
is to first pick those vertices with the highest
maximum gap.  While the algorithm is progressing, it dynamically
updates the gaps for every vertex in each step.
We now state our main result.

\begin{theorem}
\label{the:detalgo}
  The three domatic number problem can be solved by a deterministic algorithm
  running in time~$\tilde{\bigoh}(2.9416^n)$.
\end{theorem}

\begin{proofs}
Let $G = (V,E)$ be the given graph.  The algorithm seeks to find a
partition of $V$ into three disjoint dominating sets.
Note that every vertex $v \in V$ is contained in one of these sets and is
dominated by the remaining two sets, i.e., it is adjacent to at least one of
their elements. The algorithm is described in pseudo-code in the appendix,
see Figures~\ref{fig:dnp}, \ref{fig:domatic}, \ref{fig:domatic},
\ref{fig:assign}, \ref{fig:recalc}, and~\ref{fig:handle}. Since
$\delta(G) \leq \mindegree(G) + 1$, we may assume that $\mindegree(G)
\geq 2$.

The algorithm starts by initializing the potential dominating sets
$D_1, D_2$, and $D_3$ and the auxiliary sets~$A_1$, $A_2$, and~$A_3$,
setting each to the empty set.
The initial partition thus is $\mathcal{P} =
(\emptyset, \emptyset, \emptyset, V)$ and the initial triple of
auxiliary sets is $\mathcal{A} = (\emptyset, \emptyset, \emptyset)$.

Then, the recursive function $\dominate$ is called for the first time.
It is always invoked with graph $G$, a partition $\mathcal{P} = (D_1,
D_2, D_3, R)$, and a triple $\mathcal{A} = (A_1, A_2, A_3)$ of not
necessarily disjoint auxiliary sets. $\mathcal{P}$ and $\mathcal{A}$
represent a situation in which the vertices in $V - R$ have been
assigned to 
$D_1$, $D_2$, and~$D_3$, and $v \in A_i$ means that in some previous
recursive call to function $\dominate$ the vertex $v$ has been
assigned to~$D_i$ without successfully changing $\mathcal{P}$ to
contain three dominating sets.

Function $\dominate$ starts by calling $\recalculateGaps$, which
calculates all gaps with respect to $\mathcal{P}$ and~$\mathcal{A}$.
Additionally, $\openNeighbors(v)$, $\openSets(v)$, and $\balance(v)$
are determined for every vertex $v \in V$. 
Four trivial cases can occur.

\begin{construction}
  \item {\bf Case 1:} The sets $D_1$, $D_2$, and $D_3$ are dominating
  sets of graph $G$. In this case, we are done and may add the
  remaining vertices $v \in R$ to any set~$D_i$, say to~$D_1$.

  \item {\bf Case 2:} For some vertex $v \in V$, we have $\balance(v)
        < 0$. That is, there are less vertices in $R \cap N[v]$ than
        dominating sets with $v \notin N[D_i]$. Thus, no matter how the
        vertices in $R \cap N[v]$ are assigned, $\mathcal{P}$ won't
        contain three dominating sets.  We have run into a dead-end and
        return to the previous level of the recursion.

  \item {\bf Case 3:} There exists a vertex $v \in R$ that is also a
        member of two of the auxiliary sets $A_1$, $A_2$, and
        $A_3$. Hence, vertex $v$ was previously assigned to two
        distinct sets $D_i$ and~$D_j$, $1 \leq i < j \leq 3$, but the
        recursion returned without success. We assign $v$ to the only
        possible set $D_k$ left, with $i \neq k \neq j$.

  \item {\bf Case 4:} For some vertex $v \in V$, we have $\balance(v)
        = 0$ and $|| \openSets(v) || > 0$.  That is, $v$ is a critical
        vertex, since it is not dominated by all three sets~$D_1$, $D_2$,
        and $D_3$ contained in the current~$\mathcal{P}$, and there are
        as many open neighbors as open sets left for it. Note that this
        is the case for each vertex $v$ with $\degree(v) = 2$ and
        $N[v] \cap R \not= \emptyset$, as $v$ and its two neighbors
        have to be assigned to three different dominating sets. We
        select one of the at most three vertices left in $N[v] \cap
        R$, say~$u$, and call function $\assign(G, \mathcal{P},
        \mathcal{A}, u, i)$ for all $i$ with $u \notin A_i$.
\end{construction}

Function $\handleCriticalVertex$ deals with the latter three of these
trivial cases.  After they have been ruled out, one of the remaining
vertices $v \in R$ is selected and assigned to one of the three
sets~$D_i$, under the constraint that a vertex $v \in R$ cannot be
added to $D_i$ if it is already a member of~$A_i$.  This case occurs
whenever the recursion returns
because no three dominating sets could be found with this
combination.  The recursion continues by calling $\assign(G,
\mathcal{P}, \mathcal{A}, v, i)$, which adds $v$ to~$D_i$, and then
calls $\dominate(G, \mathcal{P}, \mathcal{A})$.
If no three dominating sets are found by this choice, we remember this
by adding $v$ to the set~$A_i$. A final call to $\dominate$ is made
without assigning a vertex to one potential dominating set~$D_i$.  If
this call fails, the recursion returns to the previous level.  This
completes the description of the algorithm.  We now argue that it is
correct and estimate its running time.

To see that the algorithm works correctly, note that it outputs three
sets~$D_1$, $D_2$, and~$D_3$ only if they each are dominating sets
of~$G$.  It remains to prove that these sets are definitely found in
the recursion tree.  All drop-backs within the recursion occur when,
for the current $\mathcal{P} = (D_1, D_2, D_3, R)$, we have
$\balance(v) < 0$ for some vertex $v \in V$.  Thus,
$\mathcal{P}$ cannot be modified so as to contain a correct partition
into three dominating sets on this branch of the recursion tree.
Since the algorithm checks every possible partition of $G$ into three
sets, unless it is stopped by such a drop-back, a partition into three
dominating sets will be found, if it exists.  If the algorithm does
not find three dominating sets, it eventually terminates when
returning from the first recursive call of function $\dominate$. It
reports the failure, and thus always yields the correct output.

To estimate the running time of the algorithm, an important
observation is that the recalculation of the gaps takes no more than
quadratic time in~$n$, the number of vertices of the graph $G$. Thus,
in terms of the $\tilde{\bigoh}$-notation, the running time of the
algorithm depends solely on the number of recursive calls.  Let $T(m)$
be the number of steps of the algorithm, where $m$ is the number of
potential dominating sets left for all vertices that have not been
selected as yet. Initially, every vertex may be a member of any of the
three dominating sets to be constructed (if they exist), hence $m =
3n$.

There are two scenarios where the algorithm calls function $\dominate$
recursively. If $\handleCriticalVertex$ detects a vertex $v \in V$ as
being critical, it selects a vertex $u \in N[v] \cap R$ and calls
function $\assign$ (and thus $\dominate$) for each $i$ with $u \notin
A_i$. Since every critical vertex $v \in V$ remains critical as long
as $N[v] \cap R \not= \emptyset$, function $\handleCriticalVertex$
will be called until all vertices in $N[v] \cap R$ have been assigned
to any of~$D_1$, $D_2$, and~$D_3$.  Since $|| \openSets(v) || \leq 3$,
at most three vertices in the closed neighborhood of $v$ have not been
assigned when $v$ turns critical.  By Proposition~\ref{prop:critical},
all vertices in $N[v] \cap R$ have to be assigned to different
dominating sets.  If $|| \openNeighbors(v) || = 3$, we have at most
six combinations; if we have two open neighbors for a critical vertex,
there are at most two combinations left; and finally, for one open
neighbor $u \in N[v] \cap R$, there remains only one possible choice
to assign $u$ to one of the sets~$D_1$, $D_2$, and~$D_3$. Thus, in the
worst case, we have $T(m) \leq 6 T(m-6)$, as we will handle three
vertices for which at least two choices for dominating sets are left.
With $m = 3n$, it follows that $T(m) \leq 6^{m/6} = 6^{n/2}$, i.e.,
$T(m) = \tilde{\bigoh}(2.4495^n)$.

The only other branching into two different recursive calls happens in
the main body of function $\dominate$, when selecting a vertex $v$
with the currently highest maximum gap with respect to $\mathcal{P}$
and~$\mathcal{A}$.  Two cases might occur.  On the one hand, we might
have considered a correct dominating set $D_i$ for~$v$. If $v$ had not
been looked at so far, i.e., if $v$ is not contained in any set $A_j$,
$1 \leq j \leq 3$, $j \neq i$, we have eliminated all three possible
sets for $v$ to belong to.  Thus, in this case, $T(m) = T(m -3)$. On
the other hand, if the algorithm returns from the recursion and thus
did not make the right choice for~$v$, we have $T(m) = T(m - 1)$,
since $v$ is added to~$A_i$, and function $\dominate$ is called
without assigning any vertex.  Summing up, we have $T(m) \leq T(m-1) +
T(m-3)$.  In the second case, we have already visited vertex $v$ in a
previous stage of the algorithm and unsuccessfully tried to assign it
to some set~$D_j$, with $1 \leq j \leq 3$. There are only two
dominating sets for $v$ left. Either way, if we put~$v$ into the
correct dominating set right away or fail the first time, we have
$T(m) = T(m-2)$. Summing up both cases, we have $T(m) \leq 2
T(m-2)$. Suppose that the first and the second case occur equally
often,
i.e., the algorithm considers every vertex twice.  It then follows that
\[
T(m) \leq \frac{1}{2} (T(m-1) + T(m-3)) + \frac{1}{2} (2T(m-2))
\]
with $m = 3n$. Thus, we have $T(m) = \tilde{\bigoh}(3^n)$, and the
trivial time bound cannot be beaten. To improve this running time, we
have to make sure that the recursion tree will not reach its full
depth, i.e., not all vertices are considered by the algorithm or
function $\handleCriticalVertex$ will be called for a sufficiently
large portion of the vertices.
It is clear that the algorithm has found three dominating sets once
$\area(G) = 3n$ (recall the notions from
Definition~\ref{def:notions-for-algorithm}). By selecting the maximum
gap possible for a partition $\mathcal{P}$, we try to reach this goal
as fast as possible. For every vertex $v \in R$ that we assign to one
of the potential dominating sets $D_i$, $1 \leq i \leq 3$, we increase
$\area(G)$ by $\gap(v,i)$, and additionally we add $(\gap(v,i) - 3)$
to $\surplus(G)$.

Since the vertices of degree two are critical, they and their
neighbors can be handled in time $\tilde{\bigoh}(2.4495^n)$, as
argued above.  So assume that $\mindegree(G) \geq 3$.  Then, we have
$\maxgap(G) > 3$ at the start of the algorithm. If this condition
remains to hold for at least $3n/4$ steps, we have reached $\area(G) =
3n$, and the algorithm terminates successfully.  To make use of more
than $3n/4$ vertices, $\maxgap(G)$ has to drop below four at one point
of the computation.  We exploit the fact that up to this point, the
surplus has grown sufficiently large with respect to~$n$.  Decreasing
it will force $\maxgap(G)$ to drop below three, and 
this condition can hold only for a certain portion of the remaining
vertices until the algorithm terminates.  To see this, we now analyze
the remaining steps of the algorithm after the given graph $G$ has
reached a certain maximum gap with respect to the current
$\mathcal{P}$ and~$\mathcal{A}$.

If $\maxgap(G) = 0$, the recursion stops immediately. Either we have
already found three disjoint dominating sets (in which case we put the
remaining vertices $v \in R$ into set $D_1$ and halt), or one vertex
has not been dominated by one set $D_i$ in $\mathcal{P}$ yet. Since no
positive gaps exist for the vertices $v \in R$, $\mathcal{P}$ cannot
be modified to a valid partition into three dominating sets.  Function
$\handleCriticalVertex$ returns true immediately after detecting
$\balance(v) < 0$ for some vertex $v \in V$, and function $\dominate$
drops back one recursion level. The question is how many vertices are
left in $R$ when we reach $\maxgap(G) = 0$.

\begin{lemma}
\label{lem:mainproof}
Let $G = (V,E)$ be a graph and $\mathcal{P} = (D_1, D_2, D_3, R)$ be
a partition of $V$ as in Definition~\ref{def:notions-for-algorithm}.
Let $r = || R ||$ and $\maxgap(G) = 3$. Then, for at least $r/64$
vertices in~$R$, the algorithm will not recursively call function
$\dominate$.
\end{lemma}

\sproofof{Lemma~\ref{lem:mainproof}}
Let $\maxgap(G) = k$ with $k > 0$.  Since $\gap(v,i) \leq k$ for each
$v \in R$ and for each~$i$, $1 \leq i \leq 3$, we have $\sum_{v \in R}
\sumgap(v) \leq 3kr$.  Every vertex $v$ that is selected for a set
$D_i$ with $\gap(v,i) = k$ decreases at least $k$ gaps of the vertices
in $R - \{ v \}$ by one. Otherwise, $\handleCriticalVertex$ would
have found a critical vertex $u \in N[v]$ with $N[u] \cap R = \{ v
\}$.  Then, either $|| \openSets(u) || > 1$ (which implies
$\balance(u) < 0$ and we abort), or $|| \openSets(u) || = 1$, in which
case $v$ is added to the appropriate set $D_i$ without further
branching of function $\dominate$.  Thus, if no critical vertex is
detected, selecting a vertex $v \in R$ for some set $D_i$ decreases at
least $k$ gaps, and since $v$ does not belong to $R$ anymore,
additionally all gaps previously defined for $v$ are now undefined. So
the lowest possible rate at which the gaps are decreased is related to
the maximum gap of~$G$.

Now suppose that $\maxgap(G) = 3$ and $\sumgap(v) = 9$ for all
vertices $v \in R$. We always select a vertex $v$ with the highest
summation gap of all vertices $u \in R$ with $\maxgap(u) = 3$.  As
long as there exists a vertex $v \in R$ with $\gap(v,i) = 3$ for
all~$i$, it will be selected by the algorithm.  After calling function
$\recalculateGaps$, the number of gaps equal to three will be
decreased at least by six. If exactly three other gaps of vertices in
$R - \{v\}$ decrease by one in every step, it takes at least $r/4$
vertices until $\sumgap(v) < 9$ for all $v \in R$.  Another $1/4$ of
the $3r/4$ vertices remaining have to be selected until $\sumgap(v) <
8$. Adding $1/4$ of the $9r/16$ vertices left in~$R$, we have reached
$\maxgap(G) = 2$ with $\sumgap(v) = 6$ for all vertices $v \in R$.
This implies that every defined gap is equal to two. Summing up, we
have selected
\[
\frac{1}{4} \cdot r + \frac{1}{4} \cdot \frac{3}{4} r +
 \frac{1}{4} \cdot \frac{9}{16} r = \frac{37}{64} r
\]
vertices until $\maxgap(G) = 2$, under the constraint that a minimum
number of gaps is reduced in each step, while simultaneously trying to
reduce the maximum summation gap in the fastest possible way.  This
way we reach level $\maxgap(G) = 0$ with as few vertices left in $R$
as possible, which describes the worst case that might happen.

Analogously, we can show that $\maxgap(G)$ drops from $2$ to~$1$ after
selecting another $19r/64$ vertices.  And once we have $\maxgap(G) =
1$, it takes $7r/64$ vertices to get to $\maxgap(G) = 0$.  Now, there
are $r/64$ vertices remaining in~$R$, which do not have to be
processed recursively.
\eproofof{Lemma~\ref{lem:mainproof}}

Continuing the proof of Theorem~\ref{the:detalgo}, note that
we assumed $\mindegree(G) \geq 3$, so when the gaps are initialized
for graph~$G$, we have $\mingap(v) \geq 4$ for each
vertex $v \in V$.  Thus, more than three vertices
are dominated by the selected set $D_i$ for vertex~$v$.
As long as $\maxgap(G) > 3$ is true, $\surplus(G)$ is
increasing. The only way to lower the surplus is by adding vertices
$v$ to a set $D_i$ with $\gap(v,i) < 3$.  The surplus decreases by one
when $\gap(v,i) = 2$, and it decreases by two when $\gap(v,i)~=~1$.

Let $S = \surplus(G)$ be the surplus
collected for a partition $\mathcal{P}$ until we reach a point where
$\maxgap(G) = 3$. To make use of the most recursive calls and to even
out the surplus completely, there have to be at least $r = || R ||$
vertices remaining with
\[
0 \cdot \frac{37r}{64} + 1 \cdot \frac{19r}{64}
 + 2 \cdot \frac{7r}{64} = S,
\]
so $r \geq 64S/33$. A fraction of $1/64$ of these vertices will be
handled by the algorithm without branching into more than one
recursive call, which is at least~$S/33$.
The question is how big the surplus $S$ might grow and how many vertices
are left in $R$ before $\maxgap(G) = 3$ is reached. The lowest surplus
with as few vertices in $R$ as possible occurs if $\mindegree(G) =
\maxdegree(G) = 3$. Surplus $S$ is increased by one in each step
until we arrive at $\maxgap(G) = 3$.  When selecting a vertex $v$ of
degree~$3$ for a set~$D_i$, the gap of its neighbors $u \in N(v)$
and the gaps of the neighbors of every $u$ might be decreased. Summing
up, at most $1 + 3 + 3 \cdot 2 = 10$ vertices can have decreased their
gaps for some~$i$.  After selecting at least $n/10$ vertices for each~$i$,
we have $\mingap(G) = 3$ (in the worst case). From this point on,
we cannot be sure if the next vertex selected for some $D_i$ satisfies
$\gap(v,i) > 3$. But so far we have already collected a surplus of
$S = 3n/10$, and applying this we obtain $64n/110 \leq r \leq 7n/10$.
Thus, for at least $n/110$ vertices we never branch into two different
recursive calls.  Setting $m = 3 (109n/110)$, we obtain a running time
of~$\tilde{\bigoh}(2.9416^n)$.
\end{proofs}

\section{Graphs with Bounded Maximum Degree}
\label{sec:bounded-max-degree}

As seen in the last section, the running time of the algorithm
crucially depends on the degrees of the vertices of~$G$.  If we
restrict ourselves to graphs $G$ with bounded maximum degree (say
$\Delta = \maxdegree(G)$), we can optimize our strategy in finding
three disjoint dominating sets.  In this section, we present a simple
deterministic algorithm, which has a better running time than the
algorithm from Theorem \ref{the:detalgo}, provided that $\Delta$ is low.
By using randomization, we can further improve the running time for
graphs $G$ with low maximum degree.

Before stating the two results, note that graphs with maximum degree
two can trivially be partitioned into three dominating sets, if such a
partition exists.  Every component of such a graph is either an
isolated vertex, a path, or a cycle, and each such property can be
recognized in polynomial time.

\begin{proposition}
Let $G = (V, E)$ be a given graph with $\maxdegree(G) = 2$. There
exists a partition of the vertices of $G$ into three dominating sets
if and only if every component of $G$ is a cycle of length~$k$ such
that $3$ divides~$k$.
\end{proposition}

We use the terms from Definition~\ref{def:notions-for-algorithm} in
Section~\ref{sec:algo} to describe a snapshot within the algorithm.
For any partition $\mathcal{P} = (D_1, D_2, D_3, R)$, some vertices of
$V$ have already been assigned to the potential dominating sets $D_1$,
$D_2$, and~$D_3$, while all the remaining vertices are in~$R$. The
auxiliary sets $\mathcal{A} = (A_1, A_2, A_3)$ will not be needed in
this section.  Only connected graphs are considered, as it is possible
to treat every connected component separately, producing the desired
output within the same time bounds.

Table~\ref{tab:results} lists the running times of both the
deterministic and the random algorithm, where the maximum degree of
the input graph is bounded by~$\Delta$, $3 \leq \Delta \leq 8$. Note
that the exact deterministic algorithm from Theorem~\ref{the:detalgo}
in Section~\ref{sec:algo} beats the deterministic algorithm from
Theorem~\ref{the:det-bounded} whenever $\Delta \geq 7$.

\begin{table}[!th]
\centering
\begin{tabular}{|c|c|c|c|c|c|c|}
\hline
$\Delta$ & 3 & 4 & 5 & 6 & 7 & 8 \\ \hline
deterministic & $2.2894^n$ & $2.6591^n$ & $2.8252^n$ &
 $2.9058^n$ & $2.9473^n$ & $2.9697^n$ \\ \hline
randomized & $2^n$ & $2.3570^n$ & $2.5820^n$ &
 $2.7262^n$ & $2.8197^n$ & $2.8808^n$ \\ \hline
\end{tabular}
\caption[Summary]{\label{tab:results} Results for
 $\maxdegree(G) = k$, where $3 \leq k \leq 8$}
\end{table}

\begin{theorem}
\label{the:det-bounded}
Let $G = (V, E)$ be a graph with $\maxdegree(G) = \Delta$, where
$\Delta \geq 3$.  There exists a deterministic algorithm solving the
three domatic number problem in time
$\tilde{\bigoh}(d^{\frac{n}{\Delta}})$, where
\begin{equation}
\label{equ:d}
d = \sum\limits_{a=0}^{\Delta-2} \left[ {\Delta \choose a}
  \sum\limits_{b=1}^{\Delta-a-1} {\Delta-a \choose b} \right] .
\end{equation}
\end{theorem}

\begin{proofs}
The algorithm works as follows. We start with an arbitrary vertex $v
\in V$ and assign it to the first set~$D_1$. In each step, we first
check whether we found a partition $\mathcal{P} = (D_1, D_2, D_3, R)$
into dominating sets $D_1$, $D_2$, and $D_3$. If not, one vertex
$v \in V$ is selected that is not dominated by all three sets $D_1$,
$D_2$, and $D_3$, and additionally has a vertex $u \in N[v]$ in its
closed neighborhood that has already been added to some set~$D_i$, $1
\leq i \leq 3$. It follows that $1 \leq || \openSets(v) || \leq 2$.

If $\balance(v) < 0$, we return within the recursion. Otherwise, we
try all combinations to partition the vertices in $N[v] \cap R$,
so that after this step vertex $v$ is dominated by all three
potential dominating sets.  If no such combination leads to a valid
partition, we again return within the recursion.

Suppose now that $\balance(v) \geq 0$, $|| \openSets(v) || = 2$,
and $N[v] \cap D_1 \not= \emptyset$. To obtain three disjoint
dominating sets, at least one vertex in $N[v]$ has to be assigned
to~$D_2$, and at least one vertex in $N[v]$ has to be added
to~$D_3$. This limits our choices, especially if the degree of $v$ is
bounded by some constant~$\Delta$.

To measure the running time of the algorithm, we consider the worst
case with the most possible combinations that might yield a partition
into three dominating sets.  This occurs when only one vertex $u \in
N[v]$ has already been added to one set, i.e., $|| N[v] \cap (D_1 \cup
D_2 \cup D_3) || = 1$.  If $N[v] \cap D_1 \not = \emptyset$, then any
number between $0$ and $\Delta-2$ of vertices in $N[v] \cap R$ may be
assigned to the same set~$D_1$.  Let this number be~$a$. It follows
that from one to $\Delta-a-1$ vertices remaining in $N[v] \cap R$ are
allowed to be in the next potential dominating set~$D_2$.  This is how
Equation~\ref{equ:d} for $d$ is derived.  After assigning the last
vertices in $N[v] \cap R$ to the dominating set~$D_3$, exactly $\Delta$
vertices have been removed from~$R$.  Thus, we have a worst case
running time of $\tilde{\bigoh}(d^{\frac{n}{\Delta}})$.
Table~\ref{tab:results} lists the running time for graphs with maximum
degree from three to nine.
\end{proofs}

In the next theorem, randomization is used to speed up this
procedure. Instead of assigning all vertices in the closed
neighborhood of some vertex $v \in V$ in one step, only one
or two vertices in $N[v] \cap R$ are added to the potential
dominating sets $D_1$, $D_2$, and $D_3$. The goal is to dominate
one vertex by all three sets in one step. We will select the one
or two vertices that are missing for this goal at random.

\begin{theorem}
\label{the:rand-bounded}
Let $G = (V, E)$ be a graph with $\maxdegree(G) = \Delta$, where
$\Delta \geq 3$, and let $d$ be defined as in Equation~(\ref{equ:d})
in Theorem~\ref{the:det-bounded}.  For each constant $c > 0$, there
exists a randomized algorithm solving the three domatic number problem
with error probability at most $e^{-c}$ in time
$\tilde{\bigoh}(r^{\frac{n}{2}})$, where
\begin{equation}
\label{equ:r}
r = \frac{d}{3^{\Delta-2}}.
\end{equation} 
\end{theorem}

\begin{proofs}
Let graph $G = (V,E)$ be given with $\maxdegree(G) = k$.
As in the deterministic algorithm, we start by adding a random
vertex to the set $D_1$. In every following step, a vertex
$v \in V$ is selected with $0 < || \openSets(v) || < 3$, so it
is $N[v] \cap (D_1 \cup D_2 \cup D_3) \not= \emptyset$.
If $|| \openSets(v) || = 1$, we have $N[v] \cap D_i = \emptyset$ for
one $i$ with $1 \leq i \leq 3$. We randomly choose a vertex
$u \in N[v] \cap R$ and assign it to set $D_i$, in order that
$v$ is dominated by all three sets afterwards. If $|| \openSets(v) ||
 = 2$, we randomly select two vertices $u_1, u_2 \in R$ in the
closed neighborhood of $v$. Another random choice is made when
deciding how to distribute these two vertices among the two
potential dominating sets that have not dominated $v$ up to now.

Suppose $G$ indeed has a partition into three dominating
sets. We have to measure the error rate when making our random
choices to estimate the success probability of the algorithm.
In every step, a vertex $v \in V$ is selected with at least
one vertex $u \in N[v]$ in its closed neighborhood that has
already been added to one of the sets $D_1$, $D_2$, or $D_3$.
The highest error occurs when exactly one vertex in $N[v]$ is
not included in $R$, so we restrict our analysis to this case.
To obtain a valid partition into three dominating sets, there
are at most $d$ choices left to partition the vertices
remaining in $N[v] \cap R$. Here, $d$ is the number from
Equation \ref{equ:d}. Once we selected and assigned two
vertices from $N[v] \cap R$ at random, there are
$3^{k-2}$ possibilities left to partition the vertices
in the closed neighborhood of $v$ that are still left in $R$.
Our success rate when selecting the two vertices is
therefore $3^{k-2}/d$.

To achieve an error probability of below $e^{-c}$, the
algorithm needs to be executed more than once. The repetition
number of the algorithm equals the reciprocal of
the success rate, which explains Equation \ref{equ:r}.
Since two vertices are processed in every step, the overall
running time is $\tilde{\bigoh}(r^{\frac{n}{2}})$.
\end{proofs}

\section{Conclusion}
\label{sec:conclusion}
We have shown that the three domatic number problem can be solved
by a deterministic algorithm in time~$\tilde{\bigoh}(2.9416^n)$.
Furthermore, we presented two algorithms solving the three domatic
number problem for graphs with bounded maximum degree, improving
the above time bound for graphs with small maximum degree. Although
our running times seem to be not too big of an improvement of the
trivial~$\tilde{\bigoh}(3^n)$ bound, they are to our knowledge the
first such algorithms breaking this barrier. For $k > 3$, the decision
problem of whether $\delta(G) \geq k$ can be solved in
time~$\tilde{\bigoh}(3^n)$ by Lawler's dynamic programming
algorithm for the chromatic number, appropriately modified for the domatic
number problem.  Therefore, it would not be reasonable to use our gap
approach of Section~\ref{sec:algo} to decide if $\delta(G) \geq k$
for a graph $G$ and $k > 3$.

\medskip

\noindent
{\bf Acknowledgement.} We thank Dieter Kratsch for pointing us to
Lawler's algorithm.

\bibliographystyle{alpha}

\bibliography{/home/rothe/BIGBIB/joergbib}

\clearpage

\appendix

\section{Proof of Proposition~\ref{prop:mds-dnp}}

\begin{proofs}
  Figure~\ref{fig:ex1} shows the graphs $G$ and $H$ whose existence is
  claimed.  In this figure, the numbers $i|j$ within a vertex have the
  following meaning: $i$ indicates which dominating set~$D_i$ this vertex
  belongs to in a fixed partition into three dominating sets, and $j$
  indicates a specific choice of a minimum dominating set $S$ of the graph by
  setting $j = 1$ if and only if this vertex belongs to~$S$.

\begin{figure}[!th]
\begin{center}
\input{example1.eepic}
\end{center}
\caption[Counterexamples]{\label{fig:ex1} Graphs $G$ and $H$ for
  Proposition~\ref{prop:mds-dnp}}.
\end{figure}
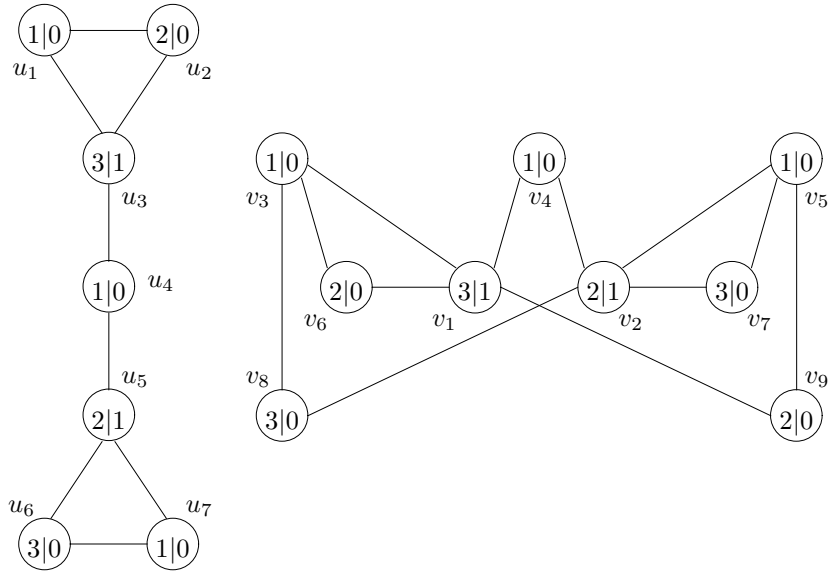
For the first assertion, look at the graph~$G$ shown on the left-hand side of
Figure~\ref{fig:ex1}.
Note that $\gamma(G) = 2$. In particular, $D = \{u_3, u_5\}$ is a minimum
dominating set of~$G$.  Note further that $\delta(G) = 3$.  In particular, a
partition into three dominating sets of $G$ is given by
$D_1 = \{u_1, u_4, u_7\}$,
$D_2 = \{u_2,u_5\}$, and
$D_3 = \{u_3,u_6\}$.
However, $D$ cannot be part of any partition into three dominating sets, since
the only neighbors of~$u_4$, namely $u_3$ and~$u_5$, belong to~$D$.

Note that the minimum dominating set $D_2 = \{u_2,u_5\}$ of $G$ defined above
indeed is part of a partition into three dominating sets.  The second part of
the proposition, however, shows that this is not always the case.  Consider
the graph $H = (V,E)$ shown on the right-hand side of Figure~\ref{fig:ex1}.
We have $\gamma(H) = 2$ by choosing the minimum dominating set $D = \{ v_1,
v_2 \}$, which is unique in this case.  Again, $\delta(H) = 3$. The only way,
up to isomorphism, to partition the vertex set of $H$ into three dominating
sets is given by
$D_1 = \{v_1, v_7, v_8\}$,
$D_2 = \{v_2, v_6, v_9\}$, and
$D_3 = \{v_3, v_4, v_5\}$. 
Thus, $\min \{||D_1||, ||D_2||, ||D_3||\} > \gamma(H)$ for each partition into
three dominating sets.
\end{proofs}

\section{Pseudo-Code of the Algorithm from Theorem~\ref{the:detalgo}}

Figures~\ref{fig:dnp}, \ref{fig:domatic}, \ref{fig:assign},
\ref{fig:recalc}, and~\ref{fig:handle} describe the algorithm from
Theorem~\ref{the:detalgo} in pseudo-code.

\begin{figure}[!ht]
\begin{construction}
\item {\bf Algorithm for the Three Domatic Number Problem}
\begin{block}
  \item {\bf Input:} Graph $G = (V,E)$ with vertex set $V = \{v_1, v_2, \ldots,
    v_n \}$  and edge set $E$
  \item {\bf Output:} Partition of $V$ into three dominating sets
   $D_1, D_2, D_3 \seq V$ or ``failure''
\end{block}
\begin{block}
    \item Set each of $D_1$, $D_2$, $D_3$, $A_1$, $A_2$,
      and $A_3$ to the empty set; 
    \item Set $R = V$;
                \item Set $\mathcal{P} = (D_1, D_2, D_3, R)$;
                \item Set $\mathcal{A} = (A_1, A_2, A_3)$;
    \item $\dominate(G,\mathcal{P}, \mathcal{A})$;
          \hfill $\slash \slash$  Start recursion
    \item {\bf output} ``failure'' and {\bf halt};
\end{block}
\end{construction}

\caption[Algorithm]{\label{fig:dnp} Algorithm for the Three Domatic
Number Problem}
\end{figure}

\begin{figure}[!ht]
\begin{construction}
\item {\bf Function} $\dominate(G,\mathcal{P},\mathcal{A})\, \{$ 
      \hfill $\slash \slash$ $\mathcal{P}$ is a partition of graph $G$, 
\item \hfill $\slash \slash$ $\mathcal{A}$ is a triple of auxiliary sets
\begin{block}
  \item $\recalculateGaps(G, \mathcal{P}, \mathcal{A})$;
  \item {\bf if} $($each $D_i$ is a dominating set$)\, \{$
  \begin{block}
    \item $D_1 = D_1 \cup R$;
    \item {\bf output} $D_1, D_2, D_3$;
  \end{block}
  \item $\}$
        \item {\bf if} $(~\mbox{not}~
               \handleCriticalVertex(G, \mathcal{P}, \mathcal{A}))\, \{$
  \begin{block}
                \item select vertex $v \in R$ with
                \begin{block}
      \item $\maxgap(v) = \maxgap(G)$ and
                        \item $\sumgap(v) = \max \{ \sumgap(u) \condition
                        u \in R \wedge \maxgap(u) = \maxgap(G)\}$;
    \end{block}
                \item find $i$ with $\gap(v,i) = \maxgap(v)$;
    \item $\assign(G, \mathcal{P}, \mathcal{A}, v, i)$;
                                \hfill $\slash \slash$ First recursive call
    \item $A_i = A_i \cup \{ v \}$;
              \hfill $\slash \slash$ If recursion fails, put $v$ in 
                                     $A_i$ and try again 
    \item $\dominate(G, \mathcal{P}, \mathcal{A})$;
                        \hfill $\slash \slash$ Second recursive call
  \end{block}
  \item $\}$
  \item {\bf return};
\end{block}
\item $\}$
\end{construction}

\caption[Algorithm]{\label{fig:domatic} Recursive function to
dominate graph $G$}
\end{figure}

\begin{figure}[!ht]
\begin{construction}
\item {\bf Function} $\assign(G, \mathcal{P}, \mathcal{A}, v, i)\, \{$ 
\begin{block}
  \item $D_i = D_i \cup \{ v \}$;
  \item $R = R - \{ v \}$;
  \item $\dominate(G, \mathcal{P}, \mathcal{A})$;
\end{block}
\item $\}$
\end{construction}

\caption{\label{fig:assign} Function to assign vertex $v$ to set $D_i$} 
\end{figure}

\begin{figure}[!ht]
\begin{construction}
\item {\bf Function} $\recalculateGaps(G, \mathcal{P}, \mathcal{A})\, \{$ 
      \hfill $\slash \slash$ $\mathcal{P}$ is a partition of graph $G$, 
\item \hfill $\slash \slash$ $\mathcal{A}$ is a triple of auxiliary sets
\begin{block}
  \item {\bf for all} $($vertices $v \in V)\, \{$
        \begin{block}
    \item {\bf if} $($vertex $v \in R)\, \{$
    \begin{block}
      \item {\bf for all} $(i = 1, 2, 3)\, \{$
      \begin{block}
        \item {\bf if} $(v \notin A_i)\, \{$
       $\gap(v,i) = || N[v] || - || \{ u \in N[v] \condition
              (\exists w \in N[u])\, [w \in D_i] \} ||$; $\}$
        \item {\bf else} $\gap(v,i) = \bot$ ;
         \hfill $\slash \slash$ $\bot$ indicates that $\gap(v,i)$ is undefined
      \end{block}
      \item $\}$
      \item $\maxgap(v) = \max_{i \in \{1,2,3\}} \{ \gap(v,i) \}$;
      \item $\mingap(v) = \min_{i \in \{1,2,3\}} \{ \gap(v,i) \}$;
      \item $\sumgap(v) = \sum_{i \in \{1,2,3\}} \gap(v,i)$;
    \end{block}
                \item $\}$
                \item $\openNeighbors(v) = \{ u \in  N[v] \condition u \in R \}$;
                \item $\openSets(v) = \{ i \in \{ 1,2,3 \} \condition
                  v \notin N[D_i] \} $;
    \item $\balance(v) = || \openNeighbors(v) || - || \openSets(v) ||$;
        \end{block}
  \item $\}$
  \item $\maxgap(G) = \max_{v \in R} \{ \maxgap(v) \}$;
  \item $\mingap(G) = \min_{v \in R} \{ \mingap(v) \}$;
\end{block}
\item $\}$
\end{construction}

\caption{\label{fig:recalc} Function to recalculate gaps
  after partition has changed} 
\end{figure}

\begin{figure}[!ht]
\begin{construction}
\item {\bf Function} boolean 
    $\handleCriticalVertex(G, \mathcal{P}, \mathcal{A})\, \{$
\begin{block}
  \item {\bf for all} $($vertices $v \in V)\, \{$
  \begin{block}
    \item {\bf if} $( \balance(v) < 0 )$ \{ \hfill $\slash \slash$
         impossible to three dominate $v$ with $\mathcal{P}$
                \begin{block}
      \item {\bf return} true;
    \end{block}
                \item \} {\bf else if}
                  $(|| \{ i \in \{ 1,2,3 \} \condition v \in A_i \} || == 2)$ \{
        \hfill $\slash \slash$ one choice for $v$ remaining
                \begin{block}
                  \item select $i$ with $v \notin A_i$;
                  \item $\assign(G, \mathcal{P}, \mathcal{A}, v, i)$;
      \item {\bf return} true;
                \end{block}
                \item \} {\bf else if}
                    $(\balance(v) == 0 ~\mbox{and}~ || \openSets(v) || > 0)$ \{
        \hfill $\slash \slash$ $v$ is critical
                \begin{block}
                  \item select $u \in N[v] \cap R$;
                        \item {\bf for all} $(i$ with $u \notin A_i$ and
                            $v$ not dominated by $D_i)$
                        \begin{block}
                          \item $\assign(G, \mathcal{P}, \mathcal{A}, u, i)$;
                        \end{block}
      \item {\bf return} true;
                \end{block}
                \item \}
  \end{block}
  \item $\}$
        \item {\bf return} false;
        \hfill $\slash \slash$ no critical vertices were found
\end{block}
\item $\}$
\end{construction}

\caption{\label{fig:handle} Function to handle the critical vertices}
\end{figure}

\end{document}

%% file: example1.eepic
\setlength{\unitlength}{0.00075489in}
\begingroup\makeatletter\ifx\SetFigFont\undefined%
\gdef\SetFigFont#1#2#3#4#5{%
  \reset@font\fontsize{#1}{#2pt}%
  \fontfamily{#3}\fontseries{#4}\fontshape{#5}%
  \selectfont}%
\fi\endgroup%
{\renewcommand{\dashlinestretch}{30}
\begin{picture}(5733,3989)(0,-10)
\put(1890,2887){\ellipse{360}{360}}
\put(1890,1087){\ellipse{360}{360}}
\put(2340,1987){\ellipse{360}{360}}
\put(3240,1987){\ellipse{360}{360}}
\put(3690,2887){\ellipse{360}{360}}
\put(5040,1987){\ellipse{360}{360}}
\put(4140,1987){\ellipse{360}{360}}
\put(5490,1087){\ellipse{360}{360}}
\put(5490,2887){\ellipse{360}{360}}
\path(2520,1987)(3060,1987)
\path(4320,1987)(4860,1987)
\path(1890,2707)(1890,1267)
\path(5490,1267)(5490,2707)
\path(2070,1087)(3960,1987)
\path(3420,1987)(5310,1087)
\path(3555,2752)(3375,2122)
\path(4005,2122)(3825,2752)
\path(5355,2752)(5175,2122)
\path(2025,2752)(2205,2122)
\path(2070,2842)(3105,2122)
\path(4275,2122)(5310,2842)
\put(1770,2797){\makebox(0,0)[lb]{\smash{{{\SetFigFont{10}{12.0}{\rmdefault}{\mddefault}{\updefault}$1|0$}}}}}
\put(1770,997){\makebox(0,0)[lb]{\smash{{{\SetFigFont{10}{12.0}{\rmdefault}{\mddefault}{\updefault}$3|0$}}}}}
\put(2220,1897){\makebox(0,0)[lb]{\smash{{{\SetFigFont{10}{12.0}{\rmdefault}{\mddefault}{\updefault}$2|0$}}}}}
\put(3570,2797){\makebox(0,0)[lb]{\smash{{{\SetFigFont{10}{12.0}{\rmdefault}{\mddefault}{\updefault}$1|0$}}}}}
\put(4920,1897){\makebox(0,0)[lb]{\smash{{{\SetFigFont{10}{12.0}{\rmdefault}{\mddefault}{\updefault}$3|0$}}}}}
\put(4020,1897){\makebox(0,0)[lb]{\smash{{{\SetFigFont{10}{12.0}{\rmdefault}{\mddefault}{\updefault}$2|1$}}}}}
\put(5370,997){\makebox(0,0)[lb]{\smash{{{\SetFigFont{10}{12.0}{\rmdefault}{\mddefault}{\updefault}$2|0$}}}}}
\put(5370,2797){\makebox(0,0)[lb]{\smash{{{\SetFigFont{10}{12.0}{\rmdefault}{\mddefault}{\updefault}$1|0$}}}}}
\put(3120,1897){\makebox(0,0)[lb]{\smash{{{\SetFigFont{10}{12.0}{\rmdefault}{\mddefault}{\updefault}$3|1$}}}}}
\put(3615,2572){\makebox(0,0)[lb]{\smash{{{\SetFigFont{10}{12.0}{\rmdefault}{\mddefault}{\updefault}$v_4$}}}}}
\put(1635,2572){\makebox(0,0)[lb]{\smash{{{\SetFigFont{10}{12.0}{\rmdefault}{\mddefault}{\updefault}$v_3$}}}}}
\put(1635,1312){\makebox(0,0)[lb]{\smash{{{\SetFigFont{10}{12.0}{\rmdefault}{\mddefault}{\updefault}$v_8$}}}}}
\put(2040,1717){\makebox(0,0)[lb]{\smash{{{\SetFigFont{10}{12.0}{\rmdefault}{\mddefault}{\updefault}$v_6$}}}}}
\put(2940,1717){\makebox(0,0)[lb]{\smash{{{\SetFigFont{10}{12.0}{\rmdefault}{\mddefault}{\updefault}$v_1$}}}}}
\put(4245,1717){\makebox(0,0)[lb]{\smash{{{\SetFigFont{10}{12.0}{\rmdefault}{\mddefault}{\updefault}$v_2$}}}}}
\put(5145,1717){\makebox(0,0)[lb]{\smash{{{\SetFigFont{10}{12.0}{\rmdefault}{\mddefault}{\updefault}$v_7$}}}}}
\put(5550,1312){\makebox(0,0)[lb]{\smash{{{\SetFigFont{10}{12.0}{\rmdefault}{\mddefault}{\updefault}$v_9$}}}}}
\put(5550,2572){\makebox(0,0)[lb]{\smash{{{\SetFigFont{10}{12.0}{\rmdefault}{\mddefault}{\updefault}$v_5$}}}}}
\put(676,2892){\ellipse{360}{360}}
\put(676,1092){\ellipse{360}{360}}
\put(225,187){\ellipse{360}{360}}
\put(1125,187){\ellipse{360}{360}}
\put(1125,3787){\ellipse{360}{360}}
\put(676,1992){\ellipse{360}{360}}
\put(225,3787){\ellipse{360}{360}}
\path(675,1807)(675,1267)
\path(675,2707)(675,2167)
\path(405,187)(945,187)
\path(405,3787)(945,3787)
\path(270,367)(630,907)
\path(720,907)(1080,367)
\path(720,3067)(1080,3607)
\path(270,3607)(630,3067)
\put(0,3472){\makebox(0,0)[lb]{\smash{{{\SetFigFont{10}{12.0}{\rmdefault}{\mddefault}{\updefault}$u_1$}}}}}
\put(1215,3472){\makebox(0,0)[lb]{\smash{{{\SetFigFont{10}{12.0}{\rmdefault}{\mddefault}{\updefault}$u_2$}}}}}
\put(765,2572){\makebox(0,0)[lb]{\smash{{{\SetFigFont{10}{12.0}{\rmdefault}{\mddefault}{\updefault}$u_3$}}}}}
\put(765,1312){\makebox(0,0)[lb]{\smash{{{\SetFigFont{10}{12.0}{\rmdefault}{\mddefault}{\updefault}$u_5$}}}}}
\put(945,1987){\makebox(0,0)[lb]{\smash{{{\SetFigFont{10}{12.0}{\rmdefault}{\mddefault}{\updefault}$u_4$}}}}}
\put(-30,412){\makebox(0,0)[lb]{\smash{{{\SetFigFont{10}{12.0}{\rmdefault}{\mddefault}{\updefault}$u_6$}}}}}
\put(1215,412){\makebox(0,0)[lb]{\smash{{{\SetFigFont{10}{12.0}{\rmdefault}{\mddefault}{\updefault}$u_7$}}}}}
\put(555,2797){\makebox(0,0)[lb]{\smash{{{\SetFigFont{10}{12.0}{\rmdefault}{\mddefault}{\updefault}$3|1$}}}}}
\put(555,1897){\makebox(0,0)[lb]{\smash{{{\SetFigFont{10}{12.0}{\rmdefault}{\mddefault}{\updefault}$1|0$}}}}}
\put(555,997){\makebox(0,0)[lb]{\smash{{{\SetFigFont{10}{12.0}{\rmdefault}{\mddefault}{\updefault}$2|1$}}}}}
\put(105,97){\makebox(0,0)[lb]{\smash{{{\SetFigFont{10}{12.0}{\rmdefault}{\mddefault}{\updefault}$3|0$}}}}}
\put(1005,97){\makebox(0,0)[lb]{\smash{{{\SetFigFont{10}{12.0}{\rmdefault}{\mddefault}{\updefault}$1|0$}}}}}
\put(105,3697){\makebox(0,0)[lb]{\smash{{{\SetFigFont{10}{12.0}{\rmdefault}{\mddefault}{\updefault}$1|0$}}}}}
\put(1005,3697){\makebox(0,0)[lb]{\smash{{{\SetFigFont{10}{12.0}{\rmdefault}{\mddefault}{\updefault}$2|0$}}}}}
\end{picture}
}

%% file: main.bbl
\newcommand{\etalchar}[1]{$^{#1}$}
\begin{thebibliography}{DGH{\etalchar{+}}02}

\bibitem[BK04]{bru-ker:t:local-search-threesat}
T.~Brueggemann and W.~Kern.
\newblock An improved local search algorithm for $3$-{SAT}.
\newblock Technical Report Memorandum No.~1709, University of Twenty,
  Department of Applied Mathematics, Enschede, The Netherlands, 2004.

\bibitem[Bon85]{bon:j:domatic-number-circular-arc-graphs}
M.~Bonuccelli.
\newblock Dominating sets and dominating number of circular arc graphs.
\newblock {\em Discrete Applied Mathematics}, 12:203--213, 1985.

\bibitem[CH77]{coc-hed:j:towards-theory-of-domination}
E.~Cockayne and S.~Hedetniemi.
\newblock Towards a theory of domination in graphs.
\newblock {\em Networks}, 7:247--261, 1977.

\bibitem[DGH{\etalchar{+}}02]{dan-etal:j:det-alg-for-ksat}
E.~Dantsin, A.~Goerdt, E.~Hirsch, R.~Kannan, J.~Kleinberg, C.~Papadimitriou,
  P.~Raghavan, and U.~Sch{\"{o}}ning.
\newblock A deterministic $(2-2/(k+1))^n$ algorithm for $k$-{SAT} based on
  local search.
\newblock {\em Theoretical Computer Science}, 289(1):69--83, October 2002.

\bibitem[Epp01a]{epp:c:algs-for-three-sat-threecolor}
D.~Eppstein.
\newblock Improved algorithms for $3$-coloring, $3$-edge-coloring, and
  constraint satisfaction.
\newblock In {\em Proceedings of the {\it 12th ACM-SIAM Symposium on Discrete
  Algorithms}}, pages 329--337. Society for Industrial and Applied Mathematics,
  2001.

\bibitem[Epp01b]{epp:c:exact-alg-for-graph-coloring}
D.~Eppstein.
\newblock Small maximal independent sets and faster exact graph coloring.
\newblock In {\em Proceedings of the {\it 7th Workshop on Algorithms and Data
  Structures}}, pages 462--470. Springer-Verlag {\it Lecture Notes in Computer
  Science \#2125}, 2001.

\bibitem[Far84]{far:j:domatic-number-strongly-chordal-graphs}
M.~Farber.
\newblock Domination, independent domination, and duality in strongly chordal
  graphs.
\newblock {\em Discrete Applied Mathematics}, 7:115--130, 1984.

\bibitem[FHK00]{fei-hal-kor:c:approximating-domatic-number}
U.~Feige, M.~Halld{\'o}rsson, and G.~Kortsarz.
\newblock Approximating the domatic number.
\newblock In {\em Proceedings of the 32nd ACM Symposium on Theory of
  Computing}, pages 134--143. ACM Press, May 2000.

\bibitem[FKW04]{fom-kra-woe:c:exact-algorithm-for-dominating-set}
F.~Fomin, D.~Kratsch, and G.~Woeginger.
\newblock Exact (exponential) algorithms for the dominating set problem.
\newblock In {\em Proceedings of the 30th International Workshop on
  Graph-Theoretic Concepts in Computer Science (WG~2004)}, pages 245--256.
  Springer-Verlag {\it Lecture Notes in Computer Science \#3353}, 2004.

\bibitem[GJ79]{gar-joh:b:int}
M.~Garey and D.~Johnson.
\newblock {\em Computers and Intractability: A Guide to the Theory of
  NP-Completeness}.
\newblock {W. H. Freeman and Company}, New York, 1979.

\bibitem[HSSW02]{hof-sch-sch-wat:c:probabilistic-threesat-alg-further-improved}
T.~Hofmeister, U.~Schöning, R.~Schuler, and O.~Watanabe.
\newblock A probabilistic 3-{SAT} algorithm further improved.
\newblock In {\em Proceedings of the 19th Annual Symposium on Theoretical
  Aspects of Computer Science}, pages 192--202. Springer-Verlag {\it Lecture
  Notes in Computer Science \#2285}, 2002.

\bibitem[HT98]{heg-tel:j:generalized-dominating-sets}
P.~Heggernes and J.~Telle.
\newblock Partitioning graphs into generalized dominating sets.
\newblock {\em Nordic Journal of Computing}, 5(2):128--142, 1998.

\bibitem[IT03]{iwa-tak:t:improved-upper-bound-threesat}
K.~Iwama and S.~Tamaki.
\newblock Improved upper bounds for $3$-{SAT}.
\newblock Technical Report TR03-053, Electronic Colloquium on Computational
  Complexity, July 2003.
\newblock 3 pages.

\bibitem[KS94]{kap-sha:j:domatic-number}
H.~Kaplan and R.~Shamir.
\newblock The domatic number problem on some perfect graph families.
\newblock {\em Information Processing Letters}, 49(1):51--56, January 1994.

\bibitem[Law76]{law:j:complexity-chromatic}
Eugene~L. Lawler.
\newblock A note on the complexity of the chromatic number problem.
\newblock {\em Information Processing Letters}, 5(3):66--67, 1976.

\bibitem[PPSZ98]{pat-pud-sak-zan:c:exptime-algorithm-for-ksat}
R.~Paturi, P.~Pudl{\'{a}}k, M.~Saks, and F.~Zane.
\newblock An improved exponential-time algorithm for $k$-{SAT}.
\newblock In {\em Proceedings of the 39th IEEE Symposium on Foundations of
  Computer Science}, pages 628--637. IEEE Computer Society Press, November
  1998.

\bibitem[Rob86]{rob:j:exact-alg-for-independent-set}
J.~Robson.
\newblock Algorithms for maximum independent sets.
\newblock {\em Journal of Algorithms}, 7:425--440, December 1986.

\bibitem[RR04]{rie-rot:j:exact-dnp}
T.~Riege and J.~Rothe.
\newblock Complexity of the exact domatic number problem and of the exact
  conveyor flow shop problem.
\newblock {\em Theory of Computing Systems}, December 2004.
\newblock On-line publication {DOI} 10.1007/s00224-004-1209-8. Paper
  publication to appear.

\bibitem[Sch02]{sch:j:constraint-satisfaction}
U.~Sch{\"{o}}ning.
\newblock A probabilistic algorithm for $k$-{SAT} based on limited local search
  and restart.
\newblock {\em Algorithmica}, 32(4):615--623, 2002.

\bibitem[Sch05]{sch:c:algorithmics-in-exponential-time}
U.~Sch{\"{o}}ning.
\newblock Algorithmics in exponential time.
\newblock In {\em Proceedings of the 22nd Annual Symposium on Theoretical
  Aspects of Computer Science}, pages 36--43. Springer-Verlag {\it Lecture
  Notes in Computer Science \#3404}, 2005.

\bibitem[Woe03]{woe:b:exact-algorithms-for-np-hard-problems}
G.~Woeginger.
\newblock Exact algorithms for {NP}-hard problems.
\newblock In M.~J{\"{u}}nger, G.~Reinelt, and G.~Rinaldi, editors, {\em
  Combinatorical Optimization: ``Eureka, you shrink!''}, pages 185--207.
  Springer-Verlag {\it Lecture Notes in Computer Science \#2570}, 2003.

\end{thebibliography}
